\documentclass[aip,reprint,amssymb,amsmath,showpacs,superscriptaddress,longbibliography]{revtex4-1}

\usepackage{graphicx}
\usepackage{dcolumn}
\usepackage{bm,dsfont}
\usepackage{epsfig}
\usepackage{epstopdf}
\usepackage[T1]{fontenc}
\usepackage{lmodern}
\usepackage{enumitem}
\usepackage{hyperref}
\usepackage{dcolumn}

\usepackage{xcolor}

\usepackage[caption=false]{subfig}

\begin{document}
\title{Competing local and global interactions in social dynamics: how important is the friendship network?}
\author{Arkadiusz J\k{e}drzejewski} 
\affiliation{Department of Operations Research and Business Intelligence, 50-370 Wroc\l{}aw University of Science and Technology, Wroc\l{}aw, Poland}
\author{Bart\l{}omiej Nowak} 
\affiliation{Department of Theoretical Physics, Wroc\l{}aw University of Science and Technology, 50-370 Wroc\l{}aw, Poland}
\author{Angelika Abramiuk} 
\affiliation{Department of Applied Mathematics, Wroc\l{}aw University of Science and Technology, 50-370 Wroc\l{}aw, Poland}
\author{Katarzyna Sznajd-Weron}
\email{katarzyna.weron@pwr.edu.pl} 
\affiliation{Department of Theoretical Physics, Wroc\l{}aw University of Science and Technology, 50-370 Wroc\l{}aw, Poland}
\date{\today}

\begin{abstract}
	Motivated by the empirical study that identifies a correlation between particular social responses and different interaction ranges, we study the $q$-voter model with various combinations of local and global sources of conformity and anticonformity. 
	The models  are investigated by means of the pair approximation and Monte Carlo simulations on Watts-Strogatz and Barab\'{a}si-Albert networks. 
	We show that  within the model with local conformity and global anticonformity, the agreement in the system is the most difficult to achieve, and the role of the network structure is the most significant.
	Interestingly, the model with swapped interaction ranges, namely with global conformity and local anticonformity, becomes almost insensitive to the changes in the network structure. The obtained results may have far reaching consequences for marketing strategies conducted via social media channels.
\end{abstract}

\maketitle
\begin{quotation}
The concept of short-range and long-range interactions is well-known in statistical physics.
By analogy, the same idea can be applied to social systems and models of opinion dynamics. 
In the era of digital communication technologies, both local and global sources of social influence can drive human behavior.
These two sources, which are typically associated with people's friends' and strangers' opinions verbalized on various online forums, may induce a  different kind of social response (e.g., conformity or anticonformity).
Many empirical studies confirm this phenomenon, and our study is directly inspired by one of them. In particular, we ask a question about the importance of these sources, influencing social responses on the individual (microscopic) level, in shaping beliefs and opinions in the society (macroscopic level). To answer this question, we analyze an opinion formation model with different combinations of local and global interactions that excite different types of social response. Our analysis indicates that the friendship network structure is the most relevant if people conform locally (to friends) and anticonform globally (to strangers), which is particularly interesting from the social point of view.
\end{quotation}

\section{introduction}
As noted by Kardar and Kaufman: \textit{``The study of competing short-range and long-range interactions is relevant to a variety of problems in statistical mechanics''} \cite{Kar:Kau:83}.
Indeed, one can easily indicate a number of natural processes in which elements interact both locally and globally \cite{Gon:etal:06, Ved:07}.
However, the mutual existence of forces with different length-scales is not only limited to physical or biological systems.
In fact, more and more empirical studies are pointing out that the overall social influence results from such a composition of local and global interactions \cite{Onn:Ree:10,Lee:Hos:Tan:15,Pan:Hou:Lui:17}.
In the era of omnipresent mass media and online social networking, people's interactions are  certainly no longer restricted to physical contacts. 
Their range, in fact, extends easily even beyond geographical borders.
This rises justified questions about the significance of these interactions in shaping trends and opinions.
How does the range of the specific type of social interactions influence the agreement in the society? What is the role of the social network in shaping opinions and how this is related to the specific type and range of interactions?

Our work is directly inspired by a recent correlation study on social influence in online movie ratings \cite{Lee:Hos:Tan:15}.
Having analyzed tendencies among reviewers to conform to already existing comments, the
authors reached a conclusion that opinions expressed by friends and strangers cause different social responses.
It turned out that those shared by the friends only led to conformity in issued reviews, whereas those of strangers might also excite anticonformity depending on the movie popularity.
These findings suggest that some types of social responses may be associated with specific interaction lengths. 
Concerning a friendship network in
this particular study, local interactions with nearest neighbors manifested only conforming
nature, whereas those global ones with strangers also displayed anticonforming properties.

These observations encouraged us to check how such differentiation between interaction ranges will impact a well-known model of opinion dynamics that already incorporates these two types of social response, that is, the $q$-voter model \cite{Nyc:Szn:Cis:12}.
In the work, we compare four versions of this model with different combinations of local and global sources of conformity and anticonformity.
Although global forces have already been considered in some studies on opinion dynamics, they were introduced in the form of effective interactions in the mean-field spirit \cite{Gon:etal:06,Jav:14,And:etal:19}.
These effective fields can be interpreted as mass media since they convey the same message to all individuals
across the system.
Our global interactions have a different character because they originate from individual opinions.
The models are analyzed on different complex networks that may represent social structures. 
We determine how the range of interactions influences the agreement in such systems and how the network structure impacts the behavior of our models.
Interestingly, the network structure is almost irrelevant to the final state of the system if proper interaction lengths are mixed. 
However, if we combine the same interaction types but with the swapped ranges, the system becomes highly sensitive to structural changes.


\section{model description}
Our study concerns one of the twin models introduced in Ref.~\cite{Nyc:Szn:Cis:12}. Both of them are modifications of the nonlinear $q$-voter model \cite{Cas:Mun:Pas:09} and are directly linked to existing psychological models of social response \cite{Jed:Szn:19}.
The first one is called the $q$-voter model with anticonformity, whereas the second one the $q$-voter with independence.
We focus on the former to align with the social responses identified in the correlation study on movie ratings \cite{Lee:Hos:Tan:15}.
The latter model is closely related to the nonlinear noisy voter model \cite{Per:etal:18,Jed:Szn:19}, and it has already been analyzed on complex networks by the use of the pair approximation and Monte Carlo simulations \cite{Jed:17}. The $q$-voter model with anticonformity, which is considered herein, has been analyzed so far only at the mean-field level on the single \cite{Nyc:Szn:Cis:12} and  the double-clique topology \cite{Kru:Szw:Wer:17}. Besides two modifications mentioned above, the nonlinear $q$-voter model has been developed in many directions \cite{Mor:etal:13,Tim:Gal:15,Jav:Squ:15,Mob:15,Sie:Szw:Wer:16,Mel:Mob:Zia:16,Mel:Mob:Zia:17,Vie:Ant:18}, for a recent review see Ref.~\cite{Jed:Szn:19}.

All of the above-mentioned $q$-voter models originate from the voter model \cite{Red:19}, which has already been studied for several decades across disciplines under different names. 
We can find related archetypal models in biology \cite{Cli:Sub:73}, genetics \cite{Mor:58}, and economy \cite{Kir:93}.
In physics, the voter model plays an important role in the studies on non-equilibrium phenomena \cite{Dor:etal:01,Mob:03,Sta:Tes:Sch:08}.
Practical applications of the voter model are known as well. For example, it has been used to recreate statistical properties of  U.S. presidential elections in the model based on social influence and recurrent mobility -- the SIRM model \cite{Fer:etal:14}. Another voter-based model has been proposed to describe the  fluctuation  of  vote  share  in the same elections \cite{Mor:His:Nak:2019}.
Moreover, a generalization of the SIRM model that also covers multiparty systems has been applied to Swedish parliamentary elections \cite{Mic:Szi:18}.

\begin{figure}[t!]
	\centering
	\epsfig{file=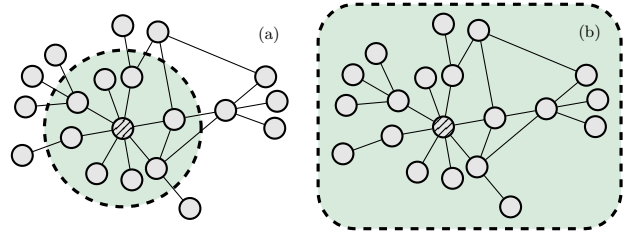,width=\columnwidth}
	\caption{\label{fig:interactions}Idea of (a) local and (b) global interactions represented for the system composed of $N=20$ voters.
		Both figures present the same friendship network. In case of local interactions, $q$ members of the influence group are chosen randomly from the nearest neighbors (encircled area) of the marked voter. In case of global interactions, the members are selected randomly from all the voters in the system. Sampling occurs without repetition.} 
\end{figure}
Opinion spreading in our setting takes place on a network that illustrates a social structure where nodes are voters, and links indicate relationships between them. 
Each node can be in two states  $j\in\{1,-1\}$, or equivalently $j\in\{\uparrow,\downarrow\}$ for simplicity of notation. The states represent different opinions, suppose a positive and a negative one. 
Voters may change them under two competing interactions recognized as different forms of social influence -- conformity and anticonformity \cite{Jed:Szn:19}.
In every time step, we choose at random a voter (a target of influence) and a group of influence comprised of $q$ randomly picked voters without repetition.
If all $q$ members of the group have the same opinion, social influence is triggered. 
Then, with probability $p$, the chosen voter acts as an anticonformist and adopts the opposite opinion to the group. Otherwise, with probability $1-p$, it behaves like a conformist and takes the group opinion.
The difference between the original model from Ref.~\cite{Nyc:Szn:Cis:12} and the one we analyze here is the set of voters form which we choose the members of the influence group.
Originally, the members of the influence group were picked from the nearest neighbors of a given voter.
In the friendship network, it translates to forces between friends -- local interactions; see Fig.~\ref{fig:interactions}(a).
In the current study, we also consider global interactions, which are not limited by the network structure.
In this case, the group members are selected from the entire population of voters, involving also strangers; see Fig.~\ref{fig:interactions}(b).

The introduction of global interactions is directly inspired by the correlation study \cite{Lee:Hos:Tan:15}, in which the differentiation between the influence of friends and strangers is made.
In this study, local interactions manifested only conforming nature, whereas global interactions could induce either conforming or anticonforming behaviors.
However, we can imagine that it is the social context that dictates the range and the type of interactions.
In fact, according to the optimal distinctiveness theory, if people feel overly similar to their group members, they try to differentiate themselves \cite{Ber:Hea:08}, so there is also a reasoning behind local anticonforming behaviors.
Therefore, we compare four versions of the $q$-voter model with all possible combinations of local and global sources of conformity and anticonformity.
We use the following bold acronyms to refer to these models:
\begin{itemize}
	\item \textbf{GAGC} -- global anticonformity and global conformity,
	\item \textbf{GALC} -- global anticonformity and local conformity,
	\item \textbf{LAGC} -- local anticonformity and global conformity,
	\item \textbf{LALC} -- local anticonformity and local conformity.
\end{itemize}
Below, we present an algorithm used for generating one step of the above models:
\begin{enumerate}
	\item \textbf{Choose randomly one node in the network.} It represents a voter that is about to reconsider its opinion under the social influence.
	\item  \textbf{Determine whether the voter will behave like a conformist or anticonformist.} With probability $p$, it acts as a conformist, otherwise, with complemantery probability $1-p$, it acts as a nonconformist.
	\item \textbf{Select the members of the influence group.} Depending on the interaction length considered in the model, choose randomly and without repetition $q$ nodes 
	\begin{enumerate}
		\item from the neighbors of the voter determined by the network structure in case of local interactions, see Fig.~\ref{fig:interactions}(a), or
		\item from all the voters in the system in case of global interactions, see Fig.~\ref{fig:interactions}(b).
	\end{enumerate}
	\item \textbf{Subject the voter to the social influence.} The voter yields to the social pressure only when all the members of the influence group have the same opinion. Thus, when the group is unanimous,
	\begin{enumerate}
		\item the voter takes the opposite opinion to the one shared by the group in case of anticonformity, or
		\item it takes the same opinion in case of conformity.
	\end{enumerate}
	If the influence group is not unanimous, the voter sticks with its old opinion.
\end{enumerate}
One Monte Carlo step corresponds to $N$ repetitions of the above steps, where $N$ is the network size (i.e., the number of voters in the system).

\section{analytical calculations}
Our analytical approach is based on the pair approximation, an enhanced version of the standard mean-field approach. This formalism is especially useful in problems involving static as well as adaptive networks \cite{Jed:Szn:19}, and it has already been applied to various binary-state dynamics \cite{Gle:13}.

Let $b$ and $c_j$ denote concentrations of active links and nodes in state $j$, respectively. 
An active link is a link that connects voters with opposite opinions.
For the notation simplicity, we put $c\equiv c_\uparrow$.
The conditional probability of choosing an active link from all the links of a node in  state $j$ is approximated by $\theta_j=b/(2c_j)$.
Having defined these quantities, the time evolution of our system can be expressed in two differential equations that in the limit of infinite network size have the following general forms \cite{Jed:17,Jed:Szn:19}:
\begin{align}
\frac{d c}{d t}=&-\sum_{j\in\{1,-1\}}c_j\sum_k P(k)\sum_{i=0}^{k}{k\choose i}\theta_j^i(1-\theta_j)^{k-i}\nonumber\\
&\times f(i,j,k)j, \label{eq:system1}\\
\frac{d b}{d t}=&\frac{2}{\langle k\rangle}\sum_{j\in\{1,-1\}}c_j\sum_k P(k)\sum_{i=0}^{k}{k \choose i}\theta_j^i(1-\theta_j)^{k-i}\nonumber\\
&\times f(i,j,k)(k-2i),
\label{eq:system2}
\end{align}
where $P(k)$ is the degree distribution of a considered network, and $\langle k\rangle$ is its average node degree.
Only function $f(i,j,k)$ is model dependent and stands for the probability that a node in state $j$ changes its opinion given that exactly $i$ out of its $k$ links are active. 
We assume that this number of active links, $i$, is binomially distributed with probability $\theta_j$.
This assumption, however, may be questionable, especially for highly clustered networks, where the neighbors of a node are likely to be connected, and thus their states may be correlated.

We consider four models with different combinations of local and global interactions. 
Each of them has its specific $f(i,j,k)$ function:
\begin{itemize}
	\item \textbf{GAGC} -- global anticonformity and global conformity:
	\begin{align}
		f(i,j,k)&=pc_j^q+(1-p)c_{-j}^q,
	\end{align}
	\item \textbf{GALC} -- global anticonformity and local conformity:
	\begin{align}
		f(i,j,k)&=pc_j^q+(1-p)\frac{i!(k-q)!}{k!(i-q)!},
	\end{align}
	\item \textbf{LAGC} -- local anticonformity and global conformity:
	\begin{align}
		f(i,j,k)&=p\frac{(k-i)!(k-q)!}{k!(k-i-q)!}+(1-p)c_{-j}^q,
	\end{align}
	\item \textbf{LALC} -- local anticonformity and local conformity:
	\begin{align}
		f(i,j,k)&=p\frac{(k-i)!(k-q)!}{k!(k-i-q)!}+(1-p)\frac{i!(k-q)!}{k!(i-q)!}.
	\end{align}
\end{itemize}
After summing over $i$ and $k$ in Eqs.~(\ref{eq:system1}) and (\ref{eq:system2}), similarly as in Ref.~\cite{Jed:17}, we get the following:
\begin{itemize}
	\item \textbf{GAGC} -- global anticonformity and global conformity:
	\begin{align}
		\label{eq:GAGCc}
		\frac{dc}{dt}=&-\sum_{j\in\{1,-1\}}c_j\left[pc_j^q+(1-p)c_{-j}^q\right]j,\\
		\label{eq:GAGCb}
		\frac{d b}{d t}=&2\sum_{j\in\{1,-1\}}c_j\left[pc_j^q+(1-p)c_{-j}^q\right](1-2\theta_j).
	\end{align}
	\item \textbf{GALC} -- global anticonformity and local conformity:
	\begin{align}
		\frac{dc}{dt}=&-\sum_{j\in\{1,-1\}}c_j\left[pc_j^q+(1-p)\theta_j^q\right]j,\\
		\frac{d b}{d t}=&\frac{2}{\langle k\rangle}\sum_{j\in\{1,-1\}}c_j\left\{pc_j^q\langle k\rangle(1-2\theta_j)\right.\nonumber\\
		&\left.+(1-p)\theta_j^q\left[\langle k\rangle-2q-2\left(\langle k\rangle-q\right)\theta_j\right]\right\}.
	\end{align}
	\item \textbf{LAGC} -- local anticonformity and global conformity:
	\begin{align}
		\frac{dc}{dt}=&-\sum_{j\in\{1,-1\}}c_j\left[p(1-\theta_j)^q+(1-p)c_{-j}^q\right]j,\\
		\frac{d b}{d t}=&\frac{2}{\langle k\rangle}\sum_{j\in\{1,-1\}}c_j\left\{p(1-\theta_j)^q\left[\langle k\rangle-2(\langle k\rangle-q)\theta_j\right]\right.\nonumber\\
		&\left.+(1-p)c_{-j}^q\langle k\rangle(1-2\theta_j)\right\}.
	\end{align}
	\item \textbf{LALC} -- local anticonformity and local conformity:
	\begin{align}
		\frac{dc}{dt}=&-\sum_{j\in\{1,-1\}}c_j\left[p(1-\theta_j)^q+(1-p)\theta_j^q\right]j,\\
		\frac{d b}{d t}=&\frac{2}{\langle k\rangle}\sum_{j\in\{1,-1\}}c_j\left\{p\left(1-\theta_j\right)^q\left[\langle k\rangle-2\left(\langle k\rangle-q\right)\theta_j\right]\right.\nonumber\\
		&\left.+(1-p)\theta_j^q\left[\langle k\rangle-2q-2\left(\langle k\rangle-q\right)\theta_j\right]\right\}.
	\end{align}
\end{itemize}
Note that after summing over $k$, we get differential equations that depend only on the average degree of a given network $\langle k\rangle$ and not on its entire node degree distribution $P(k)$. 
Of course, this result is owed to the specific form of function $f(i,j,k)$ that gives a linear function with respect to $k$ after the summation over $i$ \cite{Jed:17}.

The steady states are determined by the conditions $dc/dt=0$ and $db/dt=0$. 
Below, we present a list of formulas for the steady solutions and for the points at which the solutions at $c=1/2$ become unstable.
We designate these points as $p^*$.
For continuous phase transitions, these are the transition points between phases with (i.e., $c\neq0.5$) and without the majority opinion (i.e., $c=0.5$).
In the case of \textbf{LALC} and \textbf{LAGC} models, we were not able to obtain the explicit formulas for $b(c)$, so we present the implicit ones.

\begin{itemize}
	\item \textbf{GAGC} -- global anticonformity and global conformity:
	\\For $c\neq1/2$:
	\begin{equation}
		p=\frac{c(1-c)^q-(1-c)c^q}{(1-c)^q-c^q}
	\end{equation}
	and
	\begin{equation}
	b=2c(1-c).
	\end{equation}
	For $c=1/2$: $b=1/2$ for arbitrary values of $p$ and
	\begin{equation}
	p^*=\frac{q-1}{2q}.
	\end{equation}
	\item \textbf{GALC} -- global anticonformity and local conformity:
	\\For $c\neq1/2$:
	\begin{equation}
	p=\frac{c\theta_\uparrow^q-(1-c)\theta_{\downarrow}^q}{(1-c)^{q+1}-c^{q+1}+c\theta_\uparrow^q-(1-c)\theta_{\downarrow}^q}
	\end{equation}
	and
	\begin{widetext}
	\begin{equation}
	b=2\frac{\langle k\rangle\left[c(1-c)^{2q+1}-(1-c)c^{2q+1}\right]-q\left[c(1-c)^q+(1-c)c^q\right]\left[(1-c)^{q+1}-c^{q+1}\right]}{\langle k\rangle\left[(1-c)^{2q}-c^{2q}\right]-q\left[(1-c)^q+c^q\right]\left[(1-c)^{q+1}-c^{q+1}\right]}.
	\end{equation}
	For $c=1/2$:
	\begin{equation}
	p=\frac{\left[\langle k\rangle -2q-2(\langle k\rangle-q)b\right]b^q}{\left[\langle k\rangle -2q-2(\langle k\rangle-q)b\right]b^q-2^{-q}(1-2b)\langle k\rangle}
	\end{equation}
	\end{widetext}
	and
	\begin{equation}
	p^*=\frac{1}{\frac{1}{2^q}\frac{q+1}{q-1}\left(1+\frac{\langle k\rangle}{\langle k\rangle-q-1}\right)^q+1}.
	\end{equation}
	\item \textbf{LAGC} -- local anticonformity and global conformity:
	\\For $c\neq1/2$:
	\begin{widetext}
	\begin{equation}
	p=\frac{c(1-c)^q-(1-c)c^q}{c(1-c)^q-(1-c)c^q+(1-c)(1-\theta_{\downarrow})^q-c(1-\theta_\uparrow)^q}
	\end{equation}
	and
	\begin{align}
		0=&c(1-c)\left[(1-c)^q(1-\theta_\downarrow)^q-c^q(1-\theta_\uparrow)^q\right](1-\theta_\uparrow-\theta_\downarrow)\frac{\langle k\rangle}{q}\nonumber\\
		&+\left[c(1-c)^q-(1-c)c^q\right]\left[c\theta_\uparrow(1-\theta_\uparrow)^q+(1-c)\theta_\downarrow(1-\theta_\downarrow)^q\right].
	\end{align}
	\\For $c=1/2$:
	\begin{equation}
	p=\frac{2^{-q}(1-2b)\langle k\rangle}{2^{-q}(1-2b)\langle k\rangle-\left[\langle k\rangle -2(\langle k\rangle-q)b\right](1-b)^q}
	\end{equation}
	\end{widetext}
	and
	\begin{equation}
	p^*=\frac{1}{\frac{q+1-\Delta}{q-1+\Delta}\left(1+\frac{\Delta}{q-1}\right)^{q}+1},
	\end{equation}
	where $\Delta=\langle k\rangle-\sqrt{\langle k\rangle^2+(q-1)^2}$.
	\item \textbf{LALC} -- local anticonformity and local conformity:
	\\For $c\neq1/2$:
	\begin{widetext}
	\begin{equation}
	p=\frac{c\theta_\uparrow^q-(1-c)\theta_{\downarrow}^q}{(1-c)(1-\theta_{\downarrow})^q-c(1-\theta_\uparrow)^q+c\theta_\uparrow^q-(1-c)\theta_{\downarrow}^q}
	\end{equation}
	and
	\begin{align}
		0=&c(1-c)\left[\theta_\uparrow^q(1-\theta_\downarrow)^q-\theta_\downarrow^q(1-\theta_\uparrow)^q\right]\left(\frac{\langle k\rangle}{q}-1\right)(1-\theta_\downarrow-\theta_\uparrow)-(1-c)^2\theta_\downarrow^q(1-\theta_\downarrow)^q+c^2\theta_\uparrow^q(1-\theta_\uparrow)^q.
	\end{align}
	For $c=1/2$:
	\begin{equation}
	p=\frac{\left[\langle k\rangle -2q-2(\langle k\rangle-q)b\right]b^q}{\left[\langle k\rangle -2q-2(\langle k\rangle-q)b\right]b^q-\left[\langle k\rangle -2(\langle k\rangle-q)b\right](1-b)^q}
	\end{equation}
	\end{widetext}
	and 	
	\begin{equation}
	p^*=\frac{q-1}{\left(1+q\frac{\langle k\rangle-2}{\langle k\rangle}\right)\left(\frac{\langle k\rangle}{\langle k\rangle-2}\right)^q+q-1}.
	\end{equation}
\end{itemize}

\section{analytical result discussion}
First, let us stress that in \textbf{GAGC} dynamics, the members of the influence group are selected randomly from the entire population of voters in the system, so naturally the network topology has no impact on the model behavior. 
The system behaves exactly the same as on a complete graph.
This can be interpreted as a purely mean-field description of the $q$-voter model with anticonformity \cite{Nyc:Szn:Cis:12, Jed:Szn:19}.
Therefore, we use \textbf{GAGC} model as a benchmark for the other three models, which do involve local interactions.
\textbf{GAGC} model exhibits only continuous phase transitions for which the transition point rises along with the number of representatives in the influence group $q$, whereas the relation between steady values of $b$ and $c$ does not depend on the group size.
On the other hand, \textbf{GALC} model resembles the most the social dynamics revealed in the cited study on movie ratings \cite{Lee:Hos:Tan:15}.

\begin{figure}[!t]
	\centering
	\subfloat{\epsfig{file=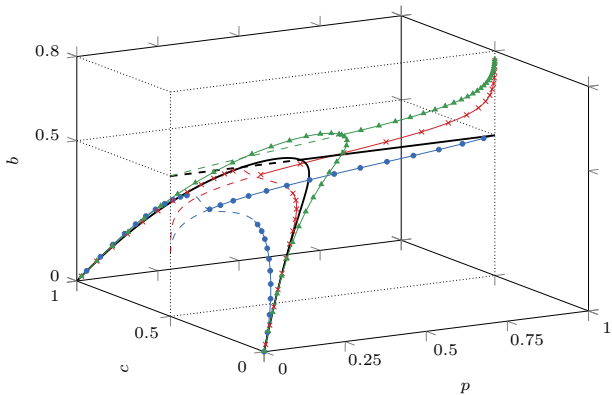,width=\columnwidth}}
	\caption{\label{fig:phase3d} Bifurcation diagram obtained within the pair approximation for $q=5$ and $\langle k\rangle=14$. Symbols correspond to different dynamics: $\bullet$ \textbf{GALC}, $\times$ \textbf{LALC}, and $\blacktriangle$ \textbf{LAGC}. Thick lines refer to the mean-field results (i.e., \textbf{GAGC} dynamics).}
\end{figure}
\begin{figure*}[!t]
	\centering
	\epsfig{file=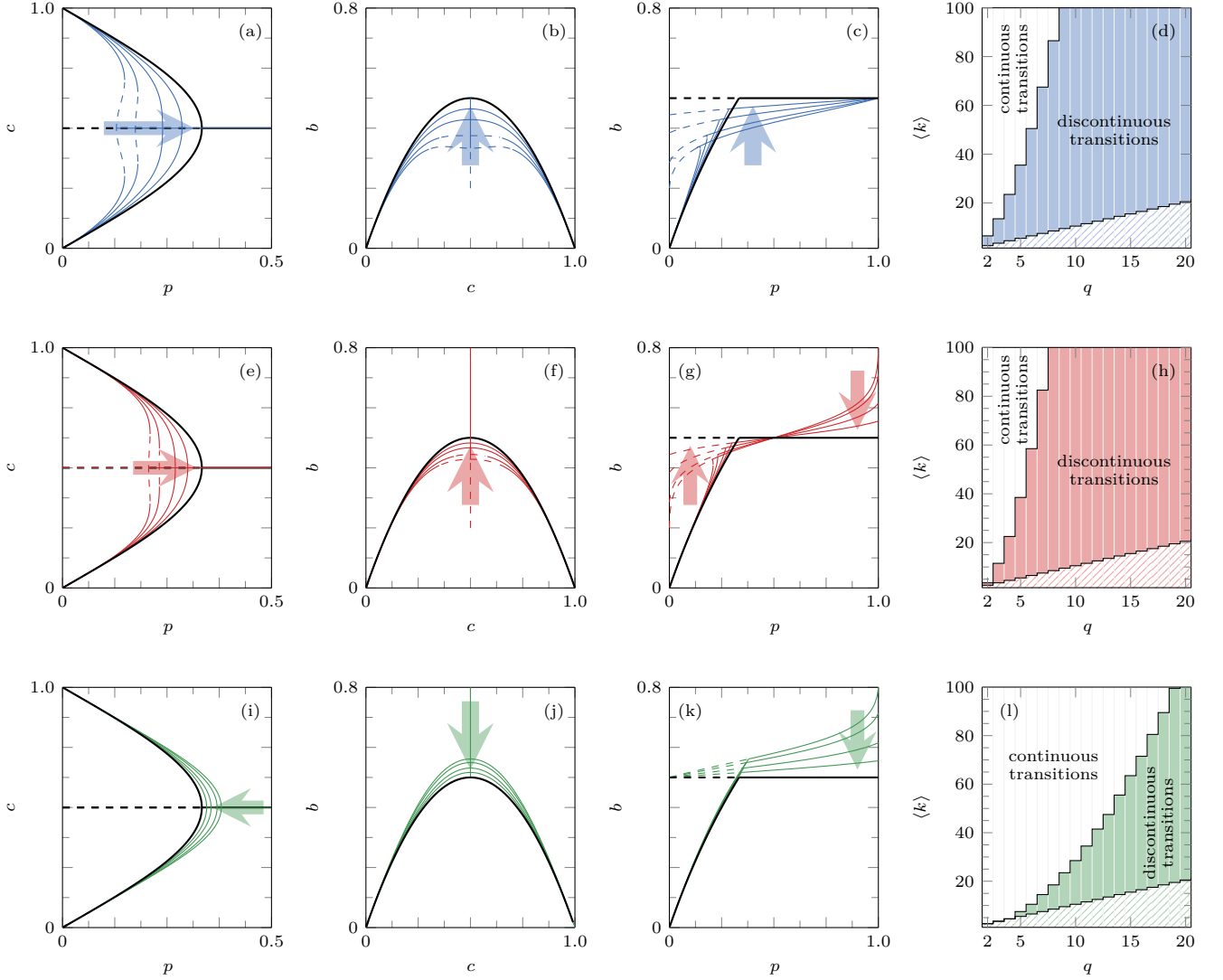,width=\textwidth}
	\caption{\label{fig:phaseDiagrams2} Response of the bifurcation diagrams to changes in the average node degree of the network $\langle k\rangle$ (the first three columns) and the phase maps that indicate the type of phase transitions predicted by the pair approximation (the last column). Figures in the same row correspond to one model: (a)-(d) \textbf{GALC}, (e)-(h) \textbf{LALC}, and (i)-(l) \textbf{LAGC}. For all the bifurcation diagrams, the group size is the same ($q=3$). Thin lines refer to different values of $\langle k\rangle\in\{8,10,16,30\}$. Arrows indicate the direction in which $\langle k\rangle$ increases. Thick lines represent \textbf{GAGC} model, which serves as a reference since all the other dynamics approach it in the limit $\langle k\rangle\rightarrow\infty$.}
\end{figure*}

Within the pair approximation, only one network parameter is essential for the behavior of the models with local interactions, namely the average node degree of the network $\langle k\rangle$.  
The formulas for the steady solutions from the previous section are presented in the bifurcation diagrams in Fig.~\ref{fig:phase3d} for sample parameters.
On the other hand, the first three columns of Fig.~\ref{fig:phaseDiagrams2} illustrate how the average node degree influences these diagrams.
One row represents one model.
The thick black lines correspond to \textbf{GAGC} dynamics as a reference.
In each subplot, the group size $q$ is fixed, and only the average degree of the network $\langle k\rangle$ changes. 
Arrows indicate the direction in which $\langle k\rangle$ increases. 

We have found that for any fixed value of $\langle k\rangle$ and $q$, the order of the dynamics (i.e., \textbf{GALC}, \textbf{LALC}, \textbf{GAGC}, and \textbf{LAGC}) from left to right in $(p,c)$ plane and from bottom to top in $(c,b)$ plane is conserved (compare subplots in Fig.~\ref{fig:phaseDiagrams2}). 
It means that the agreement is the hardest to achieve within \textbf{GALC} model, whereas the easiest within \textbf{LAGC} model.

As expected, all the dynamics approach \textbf{GAGC} model in the limit $\langle k\rangle\rightarrow\infty$.
Along this way, the change of the transition type may occur, see the last column of Fig.~\ref{fig:phaseDiagrams2}.
This limiting behavior immediately begs the question of whether our differentiation between global and local interactions makes sense in real social structures.
After all, it may turn out that the average node degree of a real friendship network is large enough to make the differences induced by these forces negligible.
In such a case, simple \textbf{GAGC} model would be sufficient.
For many real structures \cite{Alb:Bar:02}, including social networks \cite{Kos:Watt:06,Dun:etal:2015,Car:Kas:Dun:16,Zho:etal:05}, the average node degree is less than $100$ or around this value.
In fact, some studies suggest that the mean number of friends varies typically from 5 to 150 depending on the rated emotional closeness between them \cite{Dun:etal:2015,Car:Kas:Dun:16,Zho:etal:05}.
Moreover, this number remains rather stable in time even though the friendship network itself may evolve \cite{Kos:Watt:06}.
In the study on movie ratings \cite{Lee:Hos:Tan:15}, the average node degree of the network equals around $\langle k\rangle=50$.
Therefore, in the next section, we check weather there are still observable differences between the models for this reality-inspired value of $\langle k\rangle$. 
Moreover, we compare our analytical results with the outcomes of Monte Carlo simulations on Watts-Strogatz and Barab\'{a}si-Albert networks, which may represent some social structures \cite{Alb:Bar:02}.

\begin{figure*}[!ht]
	\centering
	\epsfig{file=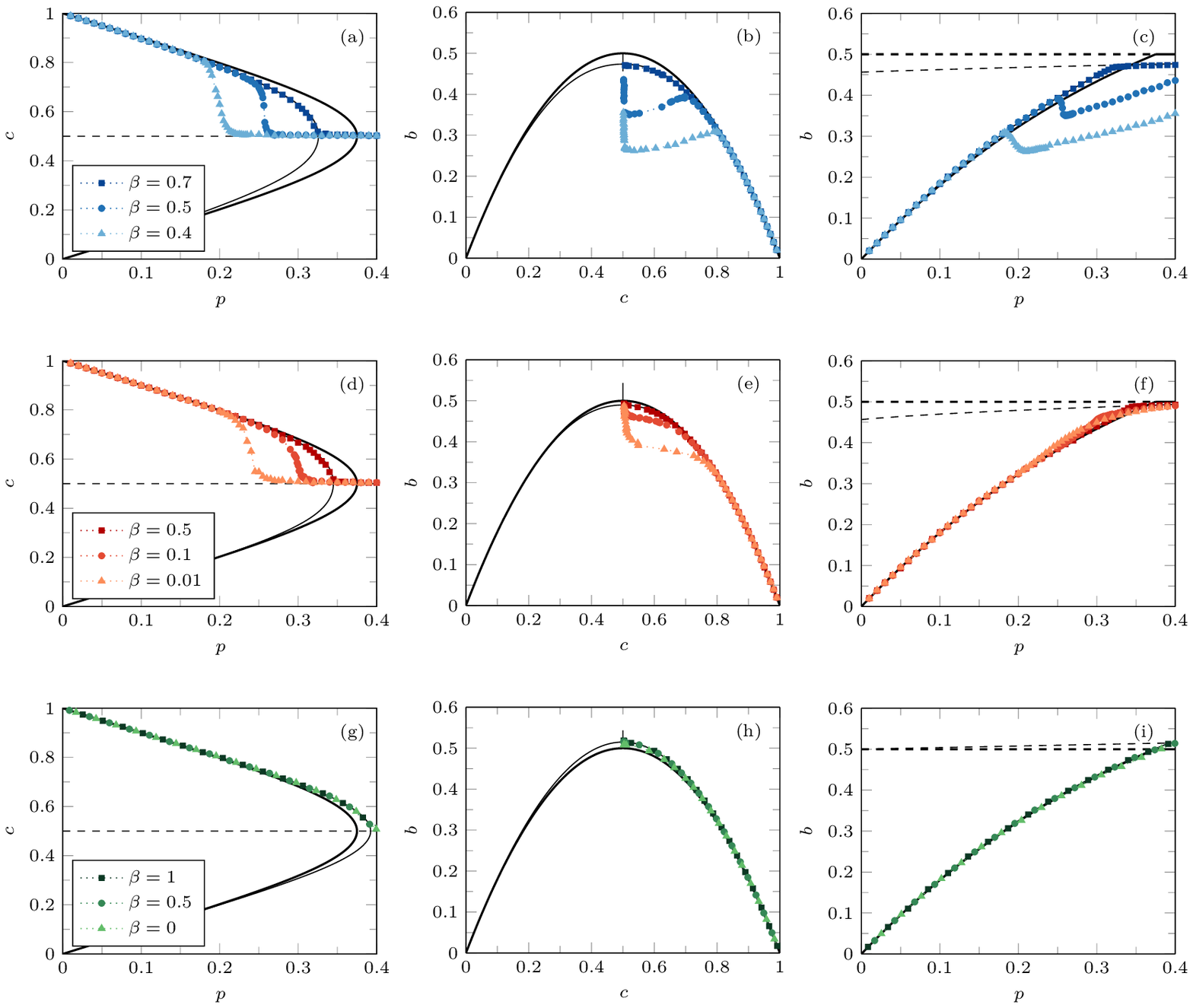,width=\textwidth}
	\caption{\label{fig:simulations}  Phase diagrams for all the models on Watts-Strogatz networks of the size $N=28160$ and the average node degree $\langle k\rangle=50$. Figures in the same row correspond to one model: (a)-(c) \textbf{GALC}, (d)-(f) \textbf{LALC}, and (g)-(i) \textbf{LAGC}. The size of a network and the average node degree correspond to the real social network used in the empirical study on movie ratings \cite{Lee:Hos:Tan:15}. The group of influence consists of $q=4$ members for all the cases. Symbols represent the outcomes of Monte Carlo simulations for different rewiring probabilities $\beta$. Thick and thin lines refer to the mean-field (i.e., \textbf{GAGC} dynamics) and the pair approximation, respectively.}
\end{figure*}

\section{monte carlo simulations}
First, we test our analytical predictions by carrying out Monte Carlo simulations on networks generated by Watts and Strogatz's algorithm \cite{Watt:Str:98}, with the link rewiring probability $\beta$.
This structure is able to recreate the small-world phenomenon present in real societies \cite{Alb:Bar:02}.
Moreover, it allows us to interpolate between regular (for $\beta=0$) and random graphs (for $\beta=1$) by tuning the parameter $\beta$.
During this interpolation, many network features change, like the average shortest path length or the clustering coefficient.  
However, the average node degree remains unchanged.
This property of the network make it particularly attractive for our study since only the average node degree of the network is relevant in the pair approximation.
In Fig.~\ref{fig:simulations}, the network size and the average node degree correspond to the real social network used in the empirical study on movie ratings \cite{Lee:Hos:Tan:15} (i.e., $N=28160$ and $\langle k\rangle=50$).
On the other hand, when it comes to the choice of the influence group size, there are some psychological evidence that groups comprised of 3 to 5 members can already achieve the maximal persuasive power \cite{Jed:Szn:19,Bon:05}.
Thus, we present the results for $q=4$.
As seen, only \textbf{LAGC} dynamics creates the same phase diagrams in the full range of the parameter $\beta$, even for highly clustered networks (small $\beta$).
The other two dynamics are more sensitive to structural changes, and they give the results consistent with the pair approximation only for high enough values of the rewiring probability.
However, graphs that describe the best real social networks are obtained for smaller values of $\beta$ (for which the network remains clustered but already has a small average path length \cite{Alb:Bar:02}).
Such graphs are called small-world networks, and in these cases \textbf{GALC} and \textbf{LALC} dynamics differ significantly from the analytical results.
The structure of a network is the most important for \textbf{GALC} model. 
In Fig.~\ref{fig:simulations}(c), we even see that there is a specific value of $\beta$ for which the dependency between $b(p)$ becomes a non-monotonic function for this dynamics.
This is a particularly interesting result if we recall that \textbf{GALC} model
corresponds to the empirical findings \cite{Lee:Hos:Tan:15}.

\begin{figure*}[!ht]
\centering
\epsfig{file=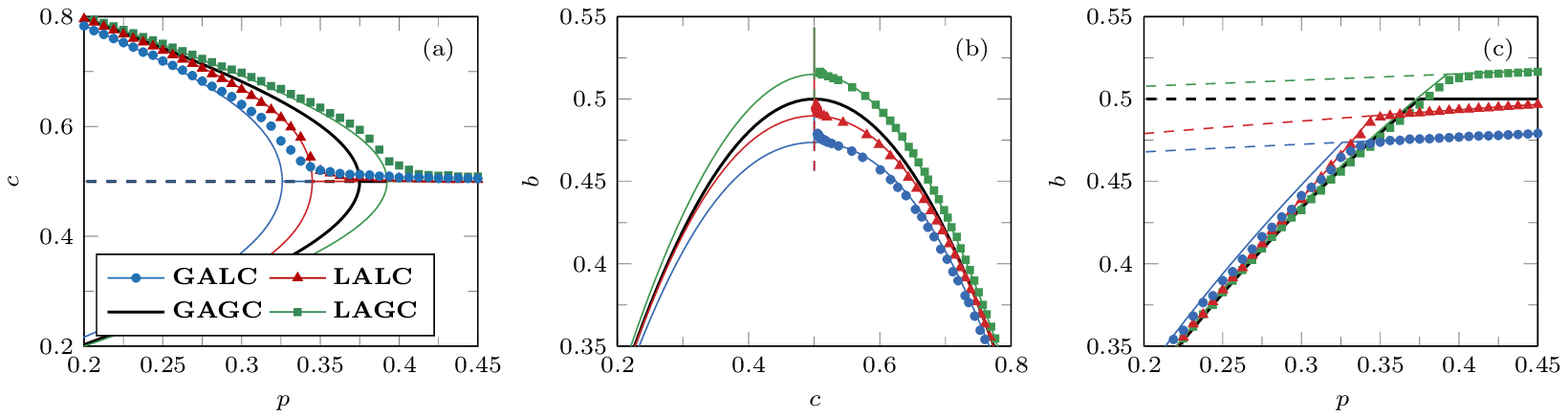,width=\textwidth}
\caption{\label{fig:simulations-BA} Phase diagrams for all the models on Barab\'{a}si-Albert networks of the size $N=28160$ and the average node degree $\langle k\rangle=50$ with relations (a) $c(p)$, (b) $b(c)$, and (c) $b(p)$. The size of a network and the average node degree correspond to the real social network used in the empirical study on movie ratings \cite{Lee:Hos:Tan:15}. The group of influence consists of $q=4$ members for all the cases. Symbols represent the outcomes of Monte Carlo simulations, whereas lines correspond to the analytical results.}
\end{figure*}

We also run the simulations on Barab\'{a}si-Albert networks, see Fig.~\ref{fig:simulations-BA}, as representatives of scale-free networks, which appear in some social structures as well, like in various collaboration networks \cite{Alb:Bar:02}.
Since these networks are much more heterogeneous with respect to degrees of nodes in comparison with Watts-Strogatz networks, one may expect the simulations to deviate more from the theoretical predictions, which rely on a certain kind of the mean-field approximation and depend only on the average node degree. 
Having compared Figs.~\ref{fig:simulations} and \ref{fig:simulations-BA}, we see, however, that the pair approximation gives more accurate predictions for heterogeneous scale-free networks than for highly clustered Watts-Strogatz networks.
We can arrive at the same conclusion studying Ref.~\cite{Jed:17}, where the $q$-voter model with independence is simulated on scale-free networks also with other power-law tails.

Lastly, since the pair approximation indicates that models with local interactions may exhibit discontinuous phase transitions, in contrast to \textbf{GAGC} dynamics, provided that the ratio of $q$ to $\langle k\rangle$ is tuned properly (see the last column of Fig.~\ref{fig:phaseDiagrams2}), we have conducted simulations with other sets of parameters. 
However, we have not detected any discontinuous phase transitions in the simulations despite their presence in the analytics. 

\section{conclusions}
Inspired by the empirical study showing association between particular types of social responses and different influence ranges, we proposed and analyzed four models of opinion formation with different combination of local and global interactions. Our analysis indicates that the agreement is the most difficult to achieve in systems with global anticonformity and local conformity (\textbf{GALC} model).
In these cases, the average opinion is the most sensitive to structural changes in the friendship network, and the range of $p$ for which one opinion dominates over the other is the smallest.
A system that exhibits such interactions is reported in the cited study on movie ratings.
In contrast, combining local anticonformity with global conformity (\textbf{LAGC} model) makes the average opinion more resistant to structural changes.
In these cases, the influence of the network structure on the final opinion is negligible for the parameters that characterize many real social systems, and only the average number of friends in the population impacts the result.
Although the limiting behavior of all the dynamics is the same,  the differences between them are significant for the typical values of the average node degree found in real-world structures.
Thus, the differentiation between interaction ranges seems to be justified. 

In the literature, there are more examples of social systems that can exhibit in-group conformity and  inter-group differentiation \cite{Ber:Hea:08}.
Such behaviors reinforce social identity. 
Therefore, some studies point to them as one of the potential factors that contribute to the recent increase in effective party polarization in the United States \cite{Iye:etal:19,Iye:Mas:Dou:19,Lu:Gao:Szy:19}. Such polarization can be observed even within a binary-vote scheme, which corresponds to our model. For example, the analysis of millions of roll-call votes cast (where a vote is coded as 1 for 'Yes' and 0 otherwise) in the U.S. Congress over the past six decades allowed to show clearly increasing political polarisation \cite{Lu:Gao:Szy:19}. 
The growth of new social media is listed as one of potential reasons for this phenomenon.

If we relate the polarisation to the disordered state in our model, i.e., the state in which both opinions are equally likely, we can conclude that the polarisation is the most supported by the short-range conformity combined with long-range anticonformity (\textbf{GALC} model), what seems to  happen in social media.

The study on social influence in online movie ratings was merely an inspiration for this study, and it cannot be used directly to validate our model. However, our study should be considered in the broader context of social polarization and its growth-promoting factors. From this point of view, obtained results seem to be realistic -- polarisation is obtained for the broadest range of control parameter $p$ within \textbf{GALC} model. Interestingly, the same version of the model is also the most sensitive to the network structure. The last result may have far reaching consequences for marketing strategies conducted via social media channels. As long as we conform to the nearest neighbors and anticonform to strangers, top-down manipulations that change the network structure can influence the public opinion significantly.

\section*{Acknowledgments}
This work has been partially supported by the National Science Center (NCN, Poland) through grants no. 2016/21/B/HS6/01256, 2016/23/N/ST2/00729 and 2018/28/T/ST2/00223. More extensive computations have been conducted using the PLGrid Infrastructure.

\section*{Data Availability Statement}
Data sharing is not applicable to this article as no new data were created or analyzed in this study.

%

\end{document}